\documentclass[lettersize,journal]{IEEEtran}
\usepackage{amsmath,amsfonts}
\usepackage{algorithmic}
\usepackage{algorithm}
\usepackage{array}
\usepackage[caption=false,font=normalsize,labelfont=sf,textfont=sf]{subfig}
\usepackage{textcomp}
\usepackage{stfloats}
\usepackage{url}
\usepackage{verbatim}
\usepackage{graphicx}
\usepackage{cite}
\usepackage{lipsum}
\usepackage{tabularx}
\usepackage{multirow}
\begin{document}

\title{Analyzing Distribution System Load Flow Through Linearization of Non-Holomorphic Functions}

\author{Ibrahim Habiballah, Wasiu Sulaimon, and Fahad  Al-Ismail,~\IEEEmembership{Member,~IEEE}

}

\markboth{IEEE TRANSACTIONS ON POWER SYSTEMS}%
{Shell \MakeLowercase{\textit{et al.}}: A Sample Article Using IEEEtran.cls for IEEE Journals}


\maketitle

\begin{abstract}
This letter presents a novel non-iterative power flow solution for radial distribution systems. In the pursuit of a linear power flow solution that seamlessly integrates into other power system operations, an approximate solution via  complex linearization of non-holomorphic functions, making no assumptions about the network's parameters was developed. This approach can be readily adapted to different load models, and its accuracy is comparable to other established conventional radial load flow analysis tools.
\end{abstract}


\section*{Nomenclature}
\addcontentsline{toc}{section}{Nomenclature}
\begin{IEEEdescription}[\IEEEusemathlabelsep\IEEEsetlabelwidth{$V_1,V_2,V_3$}]
\item[$n, m$] Number of nodes and branches.
\item[$V \in \mathbb{R}^{n}$] Node voltages.
\item[$V_{S}$] Specified voltage at the slack node.
\item[$V_{M} \in \mathbb{R}^{n - 1}$] Node voltages except the slack node.
\item[$I_{M} \in \mathbb{R}^{n - 1}$] Bus injections except the slack node. 
\item[$A \in \mathbb{R}^{m \times n}$] Incidence matrix.
\item[$1_{M} \in \mathbb{R}^{n-1}$] Unity column vector.
\item[$\left (  \cdot \right )^{\ast }$] Complex conjugate.
\item[$\odot$] Hadamard product.
\end{IEEEdescription}

\section{Introduction}
\IEEEPARstart{T}{he} objective of the load flow program is to compute the steady state of nodal voltages and angles, as well as the active and reactive power flow across all lines, based on the provided power injections at specific buses, generator outputs, and network conditions, as discussed in \cite{glover2012power}. To accomplish this, numerical techniques are frequently utilized to identify suitable operating points essential for both offline and real-time operations. Nevertheless, it's essential to note that, in general, addressing the problem in the context of both transmission and distribution system analysis is regarded as NP-hard, as highlighted in \cite{bienstock2019strong}.

Specifically within the context of distribution networks, dedicated iterative load flow techniques have been successfully developed for radial network configurations. One foundational approach for circuit analysis, which relies on the principles of Kirchhoff's Current Law (KCL) and Kirchhoff's Voltage Law (KVL), was introduced in \cite{shirmohammadi1988compensation}. Similarly, the recursive method based on the bi-quadratic equation describing the relationship between bus voltage magnitude and line flows was presented in \cite{cespedes1990new}. This latter approach has served as a benchmark for several other iterative methods. An alternative, faster, and more efficient direct approach, rooted in network topology, eliminates the necessity for both an admittance matrix and a Jacobian matrix, as outlined in \cite{teng2003direct}. This approach entirely bypasses the requirement for backward-forward sweep (BFS) methods. A comprehensive overview of all these iterative techniques can be found in \cite{martinez2011load}.

This key contribution of this letter lies in the development of an innovative, rapid, and linear load flow technique achieved through linearization of non-holomorphic functions in the complex plane. Moreover, this approach is adaptable to the ZIP model and can be seamlessly integrated into the optimization of power system operations.

\section{Methodology}

\subsection{Problem Formulation}
\begin{figure}[!t]
\centering
\includegraphics[width=2.8in]{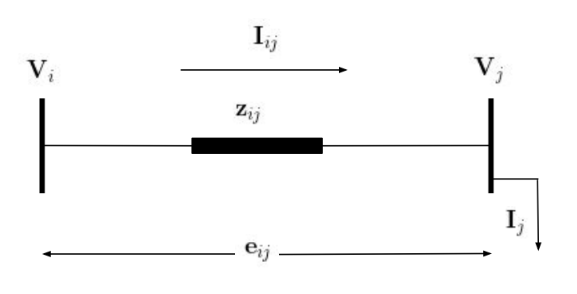}
\caption{A generic distribution system radial feeder network.}
\label{fig_1}
\end{figure}

Consider a generic radial feeder in Fig.\ref{fig_1} represented as directed graph $\mathcal{G}\left ( N,L \right )$ where $N$ and $L$ are the number of nodes and branches respectively with the property that
\begin{equation}
\label{eq1}
N = L + 1.
\end{equation}

The goal of the load flow problem is to explicitly express the nodal voltages as a function of the bus injections. In this case, we rewrite the nodal voltages, nodal current injections, and the incidence matrix as $V = \begin{bmatrix}V_{S} & V_{M}\end{bmatrix}^{T}$, $I = \begin{bmatrix}I_{S} & I_{M}\end{bmatrix}^{T}$, and $A = \begin{bmatrix}A_{S} & A_{M}\end{bmatrix}$ respectively, where $A_{S} \in \mathbb{R}^{m}$ and $A_{M} \in \mathbb{R}^{m \times n-1}$.

\begin{equation}
\label{eq2}
e = AV
\end{equation}
\begin{equation}
\label{eq3}
e = ZI_{F} 
\end{equation}
\begin{equation}
\label{eq4}
I = A^{T}I_{F}.
\end{equation}where $I_{F}$ and $e$ are the vectors of the current flows and voltage drops in the lines respectively. Exploiting (\ref{eq2})-(\ref{eq3}) by eliminating the $I_{F}$ and $e$ and with the assumption that the slack node is known with no current injections, we have
\begin{equation}
\label{eq5}
A_{M}^{-1}A_{S}V_{S} + V_{M} = A_{M}^{-1}Z\left ( A_{M}^{T} \right )^{-1}I_{M}
\end{equation} where $Z \in \mathbb{C}^{m\times m}$ is a diagonal matrix with entries as the impedance of each line. We can further simplify it by having the following result
\begin{equation}
\label{eq6}
 V_{M} - 1_{M}\cdot V_{S} = D \cdot I_{M}
\end{equation} where matrix $D$ is $A_{M}^{-1}Z\left ( A_{M}^{T} \right )^{-1}$ and $A_{M}^{-1}A_{S}V_{S}$ simplifies to a negative unity column matrix. Consequently, the $Y_{bus}$ matrix of the whole network in terms of the incidence matrix is $A^{T}CA$ which can fully expressed as 
\begin{equation}
\label{eq6a}
Y_{bus} = \begin{pmatrix}
A_{S}^TZ^{-1}A_{S} & A_{S}^{T}Z^{-1}A_{M}\\ 
A_{M}^{T}Z^{-1}A_{S}& A_{M}^TZ^{-1}A_{M}
\end{pmatrix}
\end{equation} where $C = Z^{-1}$ is a diagonal matrix of all the line admittances.

\subsection{ZIP Load Formulation}

Since the current injections are not usually specified, they can be rewritten (\ref{eq6}) as in terms of complex power injections for ZIP load models at bus $k$ as 
\begin{equation}
\label{eq7a}
I_{k} = h^{2}\cdot S_{Zk}^{\ast}\cdot V_{k} + h\cdot S_{Ik}^{\ast} + \frac{S_{Pk}^{\ast}}{V_{k}^{\ast}} 
\end{equation} where $h = 1/V_{base}$ and unity for per unit analysis. Moreover, (\ref{eq7a}) indicates that the ony source of non-linearity is introduced by the constant power load injections. Therefore, this necessitates a very accurate linearization approach for a non-iterative solution.

\subsubsection{Constant Power Load}
For the case of constant power load the expression in (\ref{eq6}) can be reformulated for each node $k$ as 
\begin{equation}
\label{eq7}
V_{k} \cdot V_{k}^{\ast} - V_{S} \cdot V_{k}^{\ast} = \sum_{j=2}^{n}D_{kj}\cdot S_{Pk}^{\ast}
\end{equation}The power flow problem in (\ref{eq7}) has non-linear term $ f(V) = \left | V_{k} \right |^{2}$ with respect to the nodal voltages. Moreover, these terms are non-holomorphic and cannot be linearized using the conventional linearization because it does not fulfil the Cauchy-Reimann condition for differentiability i.e ${f}'_{\bar{V}} \neq 0$. Using Wirtinger derivatives as discussed in \cite{garces2021mathematical}, linearization of non-holomorphic function $f\left ( V \right )$ around $V_{0}$ is given below as

\begin{equation}
\label{eq8:linear}
f\left ( V \right ) \approx f\left ( V_{0} \right ) + \left ( V-V_{0} \right ){f}'_{V}\left ( V_{0} \right ) + \left ( \bar{V}-V_{0} \right ){f}'_{\bar{V}}\left ( V_{0} \right )
\end{equation}
Then from (\ref{eq7}), the non-linear terms can be linearized around a voltage of $1+0j$ according to (\ref{eq8:linear}) as
\begin{equation}
\label{eq8}
V_{k}\cdot V_{k}^{\ast} \approx  V_{k} + V_{k}^{\ast} - 1, \quad \forall k\in \mathcal{N}
\end{equation}where $\mathcal{N}$ is the set of all nodes except the slack bus.

From (\ref{eq7}) and (\ref{eq8}) the final form for the linearized load flow can be expressed as 
\begin{equation}
\label{eq10}
\alpha V_{M}^{\ast} - V_{M} = D \cdot S_{PM}^{\ast} + 1_{M}.
\end{equation} where $\alpha$ is $V_{S} - 1$. Since $V_{S}$ is most of the time around 1.0 p.u, then the linearized form of (\ref{eq7}) can be expressed as
\begin{equation}
\label{eq11}
V_{M} = D \cdot S_{PM}^{\ast} + 1_{M} .
\end{equation}

\subsubsection{Constant Impedance Load}
In the case of constant impedance load, from (\ref{eq7a}), the expression for the node voltage at each bus and the compact form for all buses are shown below

\begin{equation}
\label{eq12}
V_{k}  - V_{S} = \sum_{j=2}^{n}D_{kj}\cdot S_{Zk}^{\ast} \cdot V_{k}
\end{equation}

\begin{equation}
\label{eq13}
\left ( 1_{M} - h^{2}\cdot D \cdot S_{ZM}^{\ast} \right ) \odot V_{M} = 1_{M} \cdot V_{S}
\end{equation}

\subsubsection{Constant Current Load}
Similarly, for the constant current load injections, the resulting formulation is shown below 

\begin{equation}
\label{eq14}
V_{k}  - V_{S} = h \cdot \sum_{j=2}^{n}D_{kj}\cdot S_{Ik}^{\ast}
\end{equation}

\begin{equation}
\label{eq15}
V_{M} = h \cdot D \cdot S_{IM}^{\ast} + 1_{M} \cdot V_{S}
\end{equation}

\subsection{ZIP Linearized Formulation}
Since the load flow formulation is rendered as a linear system, it is possible to superimpose the three formulations to form a generalized load flow equation. The compact form can be expressed below as 

\begin{equation}
\label{eq18}
V_{M}  = A^{-1}B .
\end{equation}where
\begin{align}
A & =  diag\left ( 1_{M} - h^{2}\cdot D \cdot S_{ZM}^{\ast} \right )  \\
B & = D \cdot S_{PM}^{\ast}  + h \cdot D \cdot S_{IM}^{\ast} +  1_{M} \cdot V_{S} 
\end{align}

\section{Results}
The development above is easily extensible to a three-phase unbalanced system. A simple modification is to convert phase voltages to line voltages in the case of delta connected loads similar to the transformation in \cite{garces2015linear}. The evaluation metric to be adopted  is the percentage line voltage unbalance rate as stated in \cite{pillay2001definitions} which can be expressed for a three phase node as
\begin{equation}
\label{eq18b}
\%LUVR = \frac{\left | V_{max} - V_{avg} \right |}{V_{avg}}
\end{equation}

\subsection{Case 1}
Initially, the accuracy of the proposed method is demonstrated through a comparison with established linear distribution system power flow results documented in \cite{rajagopalan1978new,cespedes1990new}. The network under examination is a balanced 30-bus radial feeder, with the source voltage set at $1.05$ p.u. To illustrate how the load flow results are influenced by the chosen linearization point in the model, the proposed method is formulated at two different linearization points: $1.05$ p.u. and $1.0$ p.u. Table~\ref{tab1} shows a decrease in accuracy in basic some basic metrics when the linearization point is established at $1.0$ p.u., which diverges significantly from the source voltage of $1.05$ p.u. The results provides additional insights into the sensitivity of the proposed technique making comparisons with other methods and underscoring the significance of selecting a linearization point in proximity to the specified source voltage.

\begin{table}
\addtolength{\tabcolsep}{-1.2pt}
\begin{center}
\label{tab:1}
\caption{Results comparison with other methods}
\label{tab1}
\begin{tabular}{| c | c | c | c | c |}
\hline
Method & $V_{S}$ (p.u.) &$V_{min}$ (p.u.)&$P_{Loss}$ (p.u.) & $Q_{Loss}$ (p.u.)\\

\hline
 \cite{rajagopalan1978new}& 1.05 & - &  0.3778 & 0.3618\\
\hline
\cite{cespedes1990new}& 1.05& 0.9147& 0.3849 & 0.3659\\ 
\hline
Proposed & \multirow{2}{*}{1.05}& \multirow{2}{*}{0.9067} & \multirow{2}{*}{0.4119} & \multirow{2}{*}{0.3911}\\ 
($V_{0} = 1.0$) & &  &  & \\ 
\hline
Proposed  & \multirow{2}{*}{1.05}& \multirow{2}{*}{0.9180} & \multirow{2}{*}{0.3513} & \multirow{2}{*}{0.3340}\\ 
($V_{0} = V_{S}$)  & &  &  & \\ 
\hline
\end{tabular}
\end{center}
\end{table}

\begin{figure}[!t]
\centering
\includegraphics[width=3.5in]{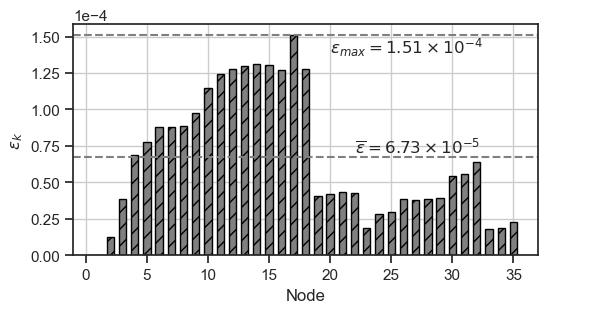}
\caption{Error between the forward-backward sweep and the proposed method}
\label{fig_2}
\end{figure}

\begin{table}
\addtolength{\tabcolsep}{-1.2pt}
\begin{center}
\caption{Simulation for balanced ZIP load model}
\label{tab2}
\begin{tabular}{| c | c | c | c | c |}
\hline
Method & $V_{S}$ (p.u.) &$V_{min}$ (p.u.)&$P_{Loss}$ (p.u.) & $Q_{Loss}$ (p.u.)\\
\hline
BFS & 1.0 & 0.9716 & 0.0293 & 0.0231 \\
\hline
\cite{garces2015linear} & 1.0 & 0.9713 & 0.0297 & 0.0234 \\
\hline
Proposed & 1.0 & 0.9717 & 0.0291 & 0.0230 \\
\hline
\end{tabular}
\end{center}
\end{table}

\begin{figure}[!t]
\centering
\includegraphics[width=3.5in]{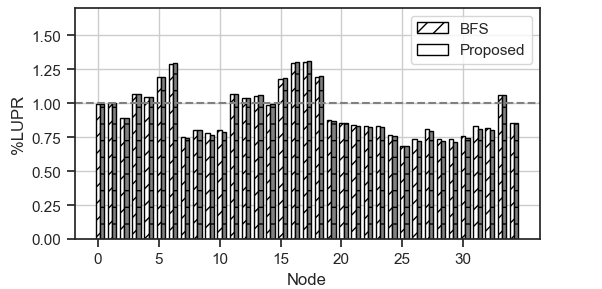}
\caption{LUVR comparison between the forward-backward sweep and the proposed method}
\label{fig_3}
\end{figure}

\subsection{Case 2}
In this scenario, the proposed method is demonstrated within the context of the ZIP model for both balanced and unbalanced case case. The data used for the load flow analysis in the balanced case is provided in \cite{garces2015linear} and all algorithms are implemented in Python environment. The evaluation metric is $\epsilon_{k}$ which is the absolute difference between the BFS algorithm and the linearized load flow at node $k$. Figure~\ref{fig_2} shows that the average error difference is very small which indicates how the proposed algorithm is significantly similar to the iterative approach. Moreover, Table \ref{tab2} further shows comparison with the iterative approach and a linear method proposed in \cite{garces2015linear} to demonstrate the accuracy of the proposed method.

For the unbalanced scenario, IEEE 37-bus test feeder data from \cite{kersting1991radial} were utilized. The system is very unbalanced and has all its load delta connected. Figure \ref{fig_3} shows the percentage LUVR for both BFS and the proposed method. The number of nodes with LUVR exceeding $1\%$ is the same for both algorithms. The $1\%$ voltage unbalance benchmark is for derating motor loads due to system unbalance.

\section{Conclusion}
An innovative, simple, and effective linear power flow model for radial distribution systems has been presented. 
The non-linearity associated with constant power loads was reformulated as a function of complex non-holomorphic terms which is then linearized with specialized complex derivatives. The accuracy of this method aligns satisfactorily with iterative techniques for both balanced and unbalanced distribution systems.

\bibliographystyle{IEEEtran}
\bibliography{IEEEabrv,Bibliography}

\vfill

\end{document}